\documentclass{article}
\usepackage{graphicx} 
\usepackage{authblk}

\title{Generalized Wilson-Cowan model with short term synaptic plasticity}

\author[1]{Tommaso Trabocchi}
\author[2]{Raffaella Burioni}
\author[3]{Lucilla de Arcangelis}
\author[1]{Duccio Fanelli}

\affil[1]{Department of Physics and Astronomy, University of Florence,  I-50019 Sesto Fiorentino, Italy}
\affil[2]{Department of Physics, University of Parma,  I-43100 Parma, Italy}
\affil[3]{Department of Mathematics and Physics, University of Campania “Luigi Vanvitelli”, I-81100 Caserta, Italy}

\begin{document}

\maketitle

\begin{abstract}
A generalized version of the Wilson-Cowan (WC) model is proposed which accounts for the evolution of the synaptic resources. Adiabatic elimination of the fast variables is performed to yield a simplified framework for the coupled interaction between active excitatory and inhibitory  neurons. The latter model is shown to smoothly converge to the benchmark WC model, when the appropriate limit is performed. Different dynamical regimes are identified for the reduced model and commented upon with reference to the original formulation of the generalized dynamics. This includes identifying limit cycle oscillations for population of available resources. 
\end{abstract}

The computational power of a neuronal system stems from the peculiar  non-linear features, as displayed by individual units, and their mutual interactions \cite{kandel:neural}.  Neurons' ability to respond to external inputs is finely orchestrated in the brain through inhibitory - excitatory patterns of intertangled interactions \cite{doi:10.1073/pnas.1712989115}. Understanding the emerging neural activity in coarse grained models which can track the average firing rate across distinct areas of the brain is essential to guide intuition and the critical handling of experimental data \cite{doi:https://doi.org/10.1002/9783527651009.ch2, RCHIALVO2004756, PhysRevLett.96.028107}. The Wilson-Cowan (WC) model \cite{WILSON19721,WILSON1973, ohira} is the reference scheme to describe the  interaction between populations of excitatory and inhibitory  neurons. According to its most elemental formulation, homogeneous populations of interconnected neurons, grouped in excitatory and inhibitory subtypes, are characterized by scalar variable that quantifies the density of cells, firing at a given time. As opposed to being detailed at the biophysical level, the system yields a sound mean field description of the overall activity of a large-scale neuronal network, by resorting to just two differential equations. The actual value of the connectivity that modulates the coupling between each population subtype and the strength of the input to each subpopulation define the key parameters of the model. By varying these latter parameters yields a wide diversity of ensuing dynamical behaviors, which can be thought as representative of the observed activity in the brain. These include multi-stability and oscillations, as well as a critical point, for the a-spatial version of the model that we shall hereafter consider \cite{PhysRevE.82.051903, ZANKOC2017504,Benayoun2010,Wallace2011}. Moreover, a bona fide critical point is also found in the limit of vanishing external field \cite{plos}. Traveling waves and self-organized patterns, are also found when explicitly accounting for the notion of space \cite{Bressloff_2012, doi:10.1073/pnas.1118672109, PhysRevE.96.062313, PhysRevE.99.012303}. 

The usual formulation of the WC model does not account for the role of synaptic plasticity, namely the ability of synapses, the connections between neurons, to modulate their strength over time.  This is a key process for both learning and memory, as it grants the brain the necessary degree of flexibility to change and adapt, in response to external stimulations. The underlying processes gather different time scales: they can be short-term (lasting milliseconds to hours) or long-term (lasting hours to years). Several mechanisms cooperate to achieve synaptic plasticity, including changes in the quantity of neurotransmitters released from a synapse or even the creation of new receptors. Starting from these premises and to fill the aforementioned gap, here we propose a straightforward modification of the WC model to account for the available synaptic resources, through the combined effect of synaptic depression and  recovery, as follows the scheme pioneered by Tsodyks and Markram (TM) \cite{doi:10.1073/pnas.94.2.719}. In the following, the model is introduced and thoroughly examined, highlighting the differences of the generalized formulation as compared to the WC benchmark setting. The phase diagram of the extended model is obtained under the assumption of an adiabatic elimination of the fast dynamical variables that quantify the amount of available synaptic resources. This yields extra non linear terms that modify the classical picture, and that fades smoothly away when the WC limit is eventually recovered. Oscillations in the fraction of available synaptic resources, a sound dynamical regime which cannot be probed under the customary WC perspective, are first predicted and then reproduced via direct numerical simulations. 

{\bf The mathematical setting} The proposed generalization of the WC model accounts for the self consistent evolution of the synaptic resources, of both excitatory and inhibitory types. The density $x_E$ (resp. $y_I$) of active excitatory (resp. inhibitory) neurons evolve according to a set of coupled mean field equations. In the following we shall set $x \equiv x_E$ and $y \equiv y_I$ for the ease of notation. The governing equations read:

\begin{eqnarray}
\label{eq1}
 \dot{x} &=& -a_E x +\left(1-x \right) \tanh \left( \omega_{EE} R_E x-\omega_{EI} R_I y+h_E \right)+I_E    \\
 \dot{y} &=& -a_I y +\left(1-y \right) \tanh \left( \omega_{IE} R_E x-\omega_{II} R_I y+h_I \right)+I_I \nonumber
\end{eqnarray}

where $\omega_{ij}$ is the strength of the synaptic connections from population $j$ to population $i$ and $R_E$  (resp. $R_I$) stands for the fraction of excitatory (resp. inhibitory) synaptic resources; $I_E$ and $I_I$ (resp.$h_E$ and $h_I$ ) are local (non local) external fields.  These latter variables are assumed to self-consistently evolve in the spirit of a TM description of the neurotransmitter release and resource replenishment process. Mathematically, we require:

\begin{eqnarray}
\label{eq2}
 \dot{R}_E &=& \frac{\left(\xi_E-R_E \right)}{\tau_{RE}}-\frac{x R_E}{\tau_{DE}} \\
\dot{R}_I &=& \frac{\left(\xi_I-R_I \right)}{\tau_{RI}}-\frac{y R_I}{\tau_{DI}} \nonumber
 \end{eqnarray}

where $\tau_{RE}$ ($\tau_{DE}$) denotes the characteristic recovery (depletion) time and $\xi_E$ represents the baseline level of non-depleted synaptic excitatory resources. Similar considerations apply to inhibitory analogue quantities, namely $\tau_{RI}$, $\tau_{DI}$, and $\xi_I$ \cite{doi:10.1073/pnas.94.2.719,PhysRevE.90.022811,PhysRevE.93.012305}. The closed set of Eqs. (\ref{eq1}) and  (\ref{eq2}) define the generalized WC model, with the inclusion of synaptic plasticity. 

We now move on to discuss how the conventional WC formulation can be effectively recovered. To this end, we posit $\xi_E=1$ and consider $\tau_{DE}>>\tau_{RE}>>1$. The non linear term in the first of the above equations (\ref{eq2}) can be hence dropped, as compared to the other, because of the assumption $\tau_{DE}>>\tau_{RE}$. On the other hand, since $\tau_{RE}$ is itself much larger than one, the density of the excitatory synaptic resources is bound to rapidly converge to the equilibrium solution, as set by the sole residual contribution of the approximated governing equation $\dot{R}_E \simeq \frac{\left(1-R_E \right)}{\tau_{RE}}$. Hence, $R_E \rightarrow 1$. Similarly,  assume $\xi_I=1$ and set  $\tau_{DI}>>\tau_{RI}>>1$. Reasoning in analogy with the above, one readily concludes that $R_I\rightarrow 1$. Working in this limit, the dynamical variables $R_E$ and $R_I$ can be replaced by their approximate equilibrium values  and the examined model turns into the canonical WC form:

\begin{eqnarray}
 \dot{x} &=& -a_E x +\left(1-x \right) \tanh \left( \omega_{EE}  x-\omega_{EI} y+h_E \right)+I_E    \\
 \dot{y} &=& -a_I y +\left(1-y \right) \tanh \left( \omega_{IE} x-\omega_{II} y+h_I \right)+I_I \nonumber
\end{eqnarray}

Having shown that the relevant WC setting can be indeed recovered in the appropriate limit $\gamma_E=\tau_{DE}/\tau_{RE}>>1$ and $\gamma_I=\tau_{DI}/\tau_{RI}>>1$, we proceed further to investigate the changes on the ensuing dynamics, as induced by the inclusion of the additional variables $R_E$ and $R_I$, for finite, though moderately large, $\gamma_E$ and $\gamma_I$. 

{\bf Adiabatic elimination} In the postulated scenario that the characteristic recovery and depletion times are large, an adiabatic elimination of the dynamical variables $R_E$ and $R_I$ can be performed. This amounts in turn to setting $\dot{R}_E =0$ (resp. $\dot{R}_I$). Solving for $R_E$ (resp. $R_I$) the stationary conditions obtained yield $R_E=\gamma_E \xi_E /(\gamma_E + x)$ and $R_I=\gamma_I \xi_I /(\gamma_I + x)$. By inserting the above conditions in  Eqs. (\ref{eq1}), one eventually gets the following closed system in the coupled variables $x$ and $y$:  
  
\begin{eqnarray}
\label{eq_ad}
 \dot{x} &=& -a_E x +\left(1-x \right) \tanh \left( \omega_{EE} \frac{\gamma_E \xi_E}{(\gamma_E + x)} x-\omega_{EI} \frac{\gamma_I \xi_I}{(\gamma_I + y)} y+h_E \right)+I_E    \\
 \dot{y} &=& -a_I y +\left(1-y \right) \tanh \left( \omega_{IE} \frac{\gamma_E \xi_E}{(\gamma_E + x)} x-\omega_{II} \frac{\gamma_I \xi_I}{(\gamma_I + y)} y+h_I \right)+I_I \nonumber
\end{eqnarray}

Interestingly enough, the WC model gets modified with the inclusion of additional non-linearities, when self-consistently accounting for the synaptic plasticity under the adiabatic approximation. These latter terms converge to one when the limit for $\gamma_E, \gamma_I \rightarrow \infty$ is performed. In this limit, and as remarked above, the original WC setting is formally recovered by further positing $\xi_E=\xi_I=1$. To reduce the number of free parameters, we will put forward the additional assumption $\gamma=\gamma_E=\gamma_I$. In the following, the landscape of possible solutions, as stemming from model (\ref{eq_ad}) for finite values of $\gamma$, will be thoroughly analyzed, when varying a subset of key reference parameters. The results obtained will be confronted to the  benchmarks setting of the limiting WC scheme. 

\begin{figure}
    \centering
    \begin{minipage}{0.4\textwidth}
    \centering
    \includegraphics[width=1\linewidth]{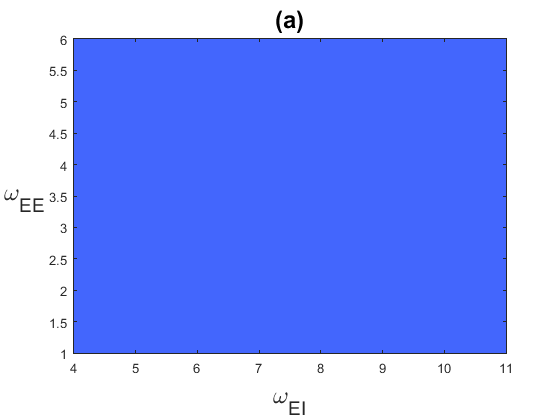}
    \end{minipage}
    \begin{minipage}{0.4\textwidth}
    \centering
    \includegraphics[width=1\linewidth]{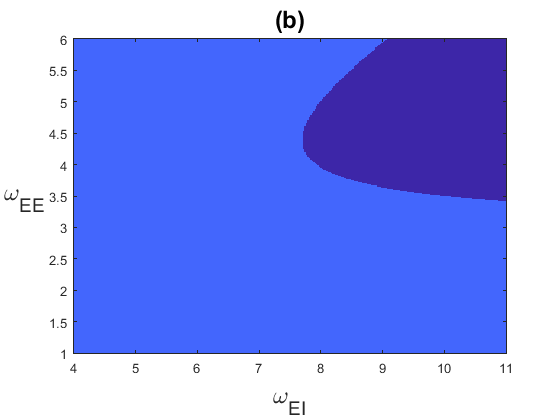}
    \end{minipage}
    \begin{minipage}{0.4\textwidth}
    \centering
    \includegraphics[width=1\linewidth]{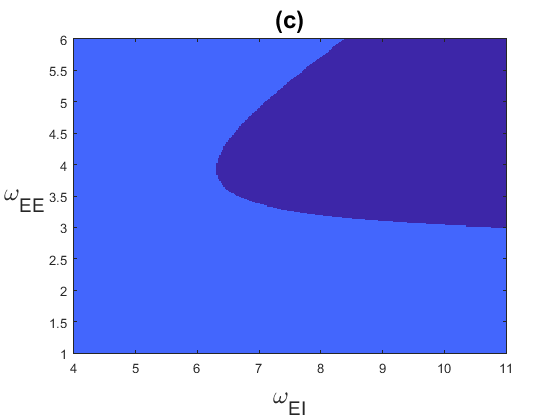}
    \end{minipage}
    \begin{minipage}{0.4\textwidth}
    \centering
    \includegraphics[width=1\linewidth]{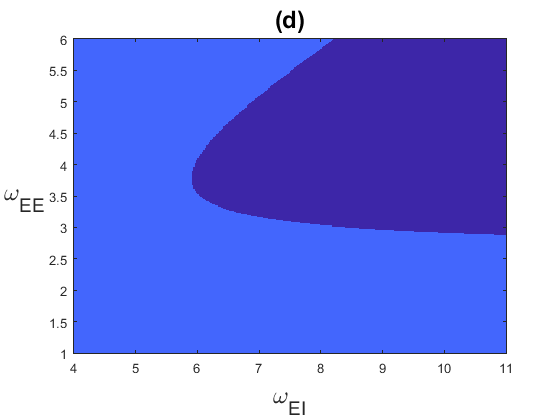}
    \end{minipage}
    \caption{The different solutions as displayed by the generalized model (\ref{eq_ad}) are illustrated in the reference plane ($\omega_{EI}$, $\omega_{EE}$), for different choices of $\gamma$. More specifically $\gamma=1$ in panel (a), $\gamma=5$ in panel (b), $\gamma=20$ in panel (c). In panel (d) the landscape of possible solutions is reported for the original WC model which is recovered, for large enough values of $\gamma$. The light blue region identifies values of 
    ($\omega_{EI}$, $\omega_{EE}$) that yields a stable fixed point solution. Working in the dark blue domain, a limit cycle is found. The other parameters are set to the following values: $a_E=a_I=1$, $h_E=-0.4$, $h_I=-0.5$, $I_E=0.5$, $I_I=0$,$\xi_E=\xi_I=1$, $\omega_{IE}=2$ and $\omega_{II}=0.1$.}
    \label{Figure1}
\end{figure}

{\bf Identifying different dynamical regimes.} Let us begin by setting $\xi_E=\xi_I=1$ (this done to eventually recover the correct WC model in the proper limit). To uncover the 
 different dynamical regimes displayed by model (\ref{eq_ad}), we scan the plane ($\omega_{IE}$, $\omega_{EE}$), for different choices of $\gamma$, while keeping the other parameters fixed. Results of the analysis are reported in Figure \ref{Figure1}. The lower right figure (panel (d)) refers to the original WC framework, which is formally recovered for diverging $\gamma$: in the light blue region, the WC model yields a stable fixed point. In the dark blue region, the fixed point turns unstable and a limit cycle solution is found. Already at $\gamma=20$, the landscape of possible solution resembles closely that obtained for the limiting WC setting. When $\gamma$ gets further reduced, the region associated to the limit cycle solution shrinks ($\gamma=5$) and eventually disappears ($\gamma=1$). This behavior is direct consequence of synaptic strength modulation introduced by short-term plasticity in Eqs. (\ref{eq2}). This adaptation can be therefore interpreted as a form of dissipation of resources which hampers limit cycle occurrence. Interestingly, the period of the displayed oscillations (data not shown) depends on the chosen value of $\gamma$. Larger $\gamma$ yields longer lasting cycles. 
 
\begin{figure}
    \centering
    \begin{minipage}{0.4\textwidth}
    \centering
    \includegraphics[width=1\linewidth]{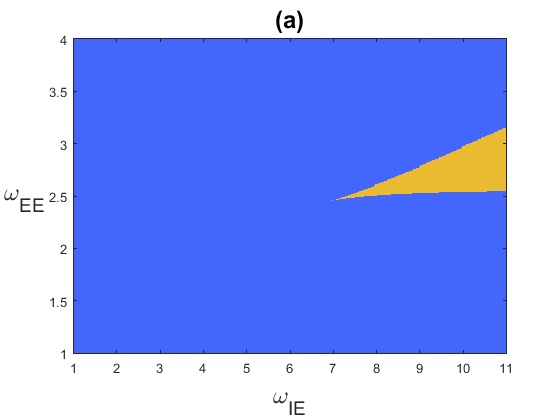}
    \end{minipage}
    \begin{minipage}{0.4\textwidth}
    \centering
    \includegraphics[width=1\linewidth]{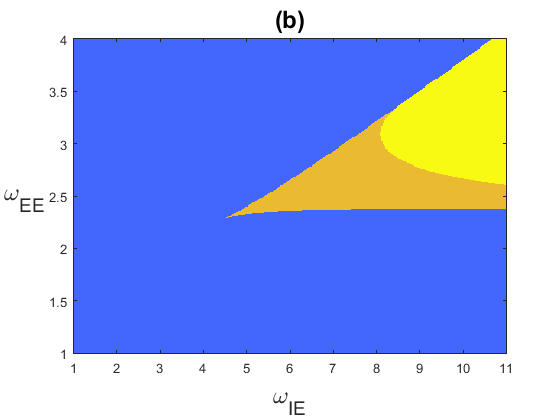}
    \end{minipage}
    \begin{minipage}{0.4\textwidth}
    \centering
    \includegraphics[width=1\linewidth]{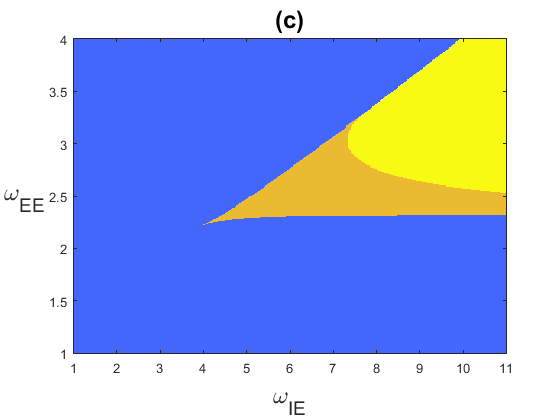}
    \end{minipage}
    \begin{minipage}{0.4\textwidth}
    \centering
    \includegraphics[width=1\linewidth]{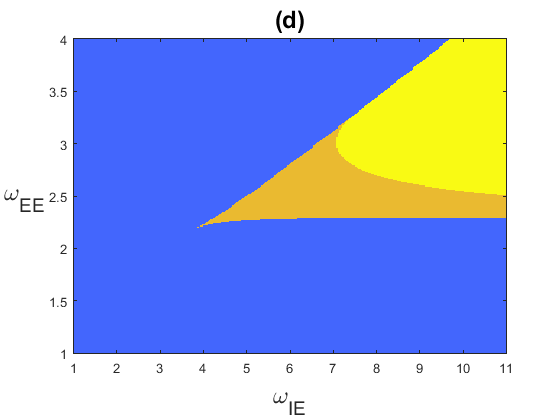}
    \end{minipage}
    \caption{The landscape of possible solutions of   model (\ref{eq_ad}) are displayed when scanning the plane ($\omega_{IE}$, $\omega_{EE}$), for a different choice of $\omega_{EI}$, as compared to Figure \ref{Figure1} (here, in particular, $\omega_{EI}=1.5$ while the other parameters stay unchanged). The region drawn in orange identifies a bistable regime (two stable fixed points, separated by a saddle point) while the region drawn in yellow contains a stable fixed point, a saddle and an unstable one . When the parameters are set to the values that correspond to the region colored with light blue, the system shows a stable fixed point. When $\gamma$ gets reduced, the region of bistability shrinks. Panel (a) corresponds to $\gamma=1$, panel (b) to $\gamma=5$ and panel (c) to $\gamma=20$. The WC solution is depicted in panel (c) and it is formally recovered for sufficiently large values of $\gamma$.}
    \label{Figure2}
\end{figure}

Complementary insight can be gained when scanning the same reference plane,  for a different choice of the residual parameters, fixed to nominal values. In Figure \ref{Figure2} the region highlighted in yellow stands for a bistable regime (two stable fixed points are found, separated by a saddle point), while in the portion of the plane colored with light blue drawing, the system converges to a stable fixed point. When reducing $\gamma$, the region of bistability gets significantly reduced. Interestingly, the position usually implemented in the linear noise approximation of the Wilson Cowan Master equation \cite{Benayoun2010}, assuming that synaptic strengths solely depend on the presynaptic population, namely $\omega_{EE}=\omega_{IE}$ and $\omega_{II}=\omega_{EI}$, is equivalent to vary parameters along the main diagonal in Fig. \ref{Figure2}.

\begin{figure}
    \centering
    \begin{minipage}{0.4\textwidth}
    \centering
    \includegraphics[width=1\linewidth]{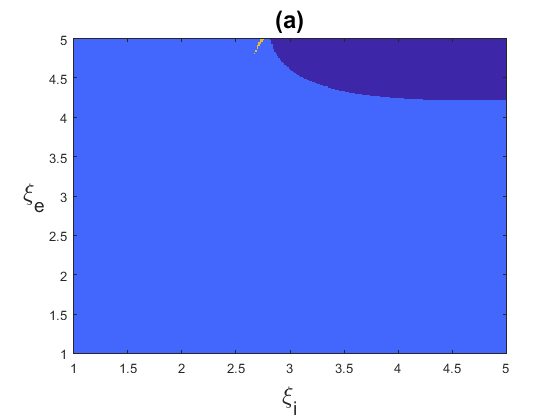}
    \end{minipage}
    \begin{minipage}{0.4\textwidth}
    \centering
    \includegraphics[width=1\linewidth]{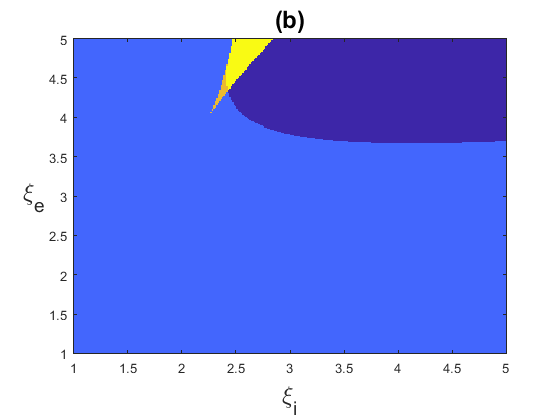}
    \end{minipage}
    \begin{minipage}{0.4\textwidth}
    \centering
    \includegraphics[width=1\linewidth]{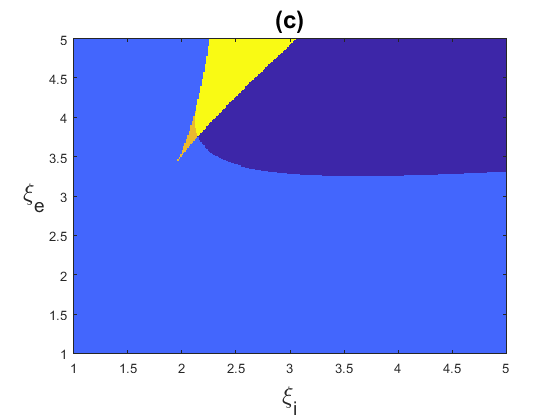}
    \end{minipage}
    \begin{minipage}{0.4\textwidth}
    \centering
    \includegraphics[width=1\linewidth]{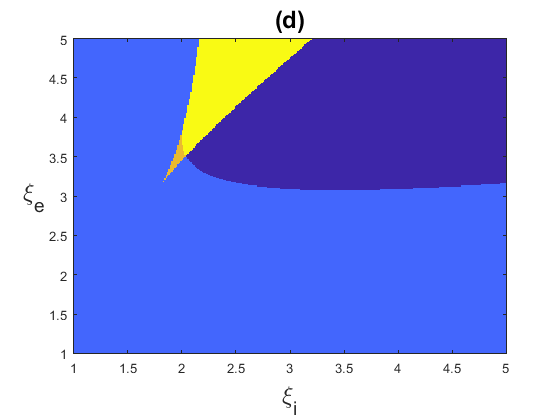}
    \end{minipage}
    \caption{Different dynamical regime for model (\ref{eq_ad}) are characterized as a function of the 
    the baseline level of non depleted synaptic resources, namely ($\xi_I$,$\xi_E$) and for different choices of the parameter $\gamma$. Panel (a) corresponds to $\gamma=1$, panel (b) to $\gamma=2$ and panel (c) to $\gamma=7$. Panel (d) refers to the WC model: in this setting the parameters $\omega_{EE}$ and $\omega_{II}$ are modulated, while keeping the ratios $\omega_{EE}/\omega_{EI}$ and $\omega_{II}/\omega_{IE}$ frozen to the values used in generating the outputs reported in panels (a),(b) and (c). The other parameters have been here set to the values $a_E=a_I=1$, $h_E=-0.4$, $h_I=-0.5$, $I_E=0.5$, $I_I=0$,$\omega_{EE}=\omega_{IE}=1$, $\omega_{EI}=1.5$ and $\omega_{II}=0.1$. For  $(\xi_I,\xi_E)$ in the light blue region, the system converges to a stable fixed point. In the dark blue region, the fixed point is unstable and the system display a limit cycle. For larger $\gamma$, two additional regions are found. The system has three fixed point  two saddle nodes and one stable attractor in the yellow region. The system is instead bistable (two stable nodes and one saddle node) in the region depicted in orange.}
    \label{Figure3}
\end{figure}

Notably, by working in the generalized framework of equations (\ref{eq_ad}) (or, equivalently, assuming the initial  setting where the dynamics of $R_x$ and $R_y$ is explicitly accounted for) one can probe the effect of modulating the baseline level of non depleted synaptic resources, namely ($\xi_I$,$\xi_E$), on the predicted dynamical regimes. In the original WC model, this condition can be artificially mimicked by simultaneously changing $\omega_{EE}$ and $\omega_{EI}$ (resp. $\omega_{EI}$ and $\omega_{II}$) over the region of interest, while forcing  their ratio to stay constant. The results of these additional investigations are reported in Figure \ref{Figure3}. 

For $(\xi_I,\xi_E)$ laying in the light blue region, the system displays a stable fixed point. On the contrary, in the dark blue region, the fixed point is unstable and the system displays a limit cycle.
By increasing $\gamma$, two more regions appear. Inside the region depicted in yellow, the system has three fixed point, two saddle nodes, and one stable attractor. In correspondence of the region represented in orange, the system is bistable (two stable nodes and one saddle node).

Summing up, by reformulating the dynamics of the 
WC model to account for synaptic plasticity, one can uncover a rich zoology of diverse dynamical regimes, which significantly extends beyond the conventional plot. 

\begin{figure}
    \centering
    \begin{minipage}{0.4\textwidth}
    \centering
    \includegraphics[width=1\linewidth]{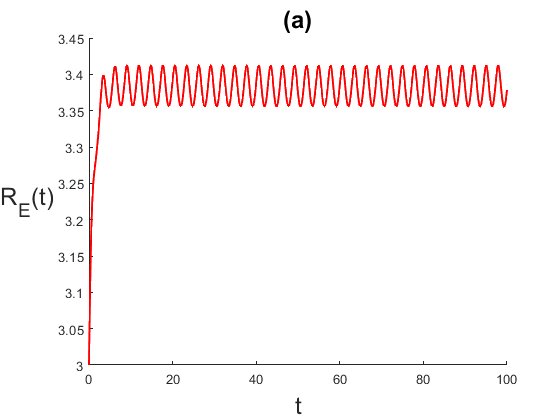}
    \end{minipage}
    \begin{minipage}{0.4\textwidth}
    \centering
    \includegraphics[width=1\linewidth]{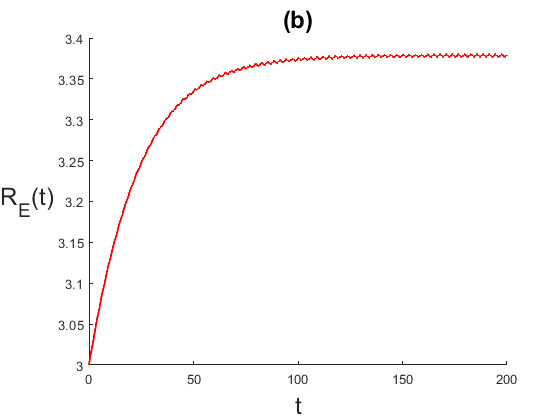}
    \end{minipage}
    \begin{minipage}{0.4\textwidth}
    \centering
    \includegraphics[width=1\linewidth]{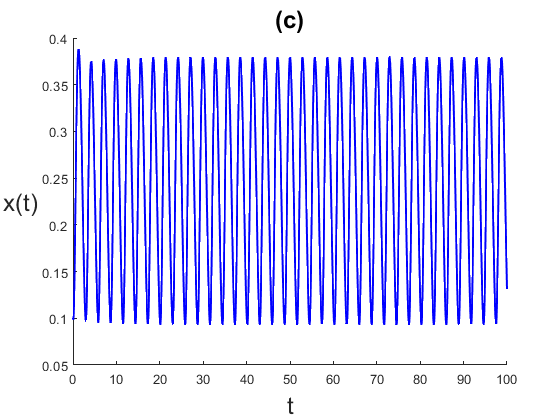}
    \end{minipage}
    \begin{minipage}{0.4\textwidth}
    \centering
    \includegraphics[width=1\linewidth]{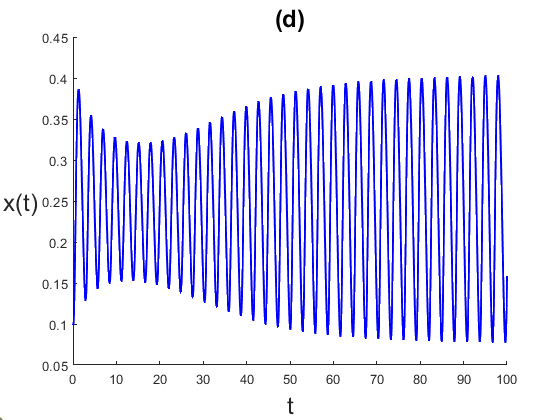}
    \end{minipage}
    \caption{Two upper panels: the resource variables ($R_I$ in blue and $R_E$ in red), oscillate in time. The predicted limit cycle for $x$ and $y$ reverberates on the coupled variables $R_E$ and $R_I$. The oscillations fades however away for sufficiently large values of $\tau_R$ (and $\tau_D$) at constant $\gamma$. Species $x$ and $y$ keep on oscillating , independently of the chosen time scales (see lower panels). Here, $a_E=a_I=1$, $h_E=-0.4$, $h_I=-0.5$, $I_E=0.5$, $I_I=0$,$\omega_{EE}=\omega_{IE}=1$, $\omega_{EI}=1.5$, $\omega_{II}=0.1$, $\xi_E=3.5$, $\xi_I=3$ and $\gamma=\frac{\tau_d}{\tau_r}=7$.}
    \label{Figure4}
\end{figure}

{\bf Testing the predictions: oscillations in the available resources}. As a representative example, we recall that the reduced model (\ref{eq_ad}) predicts the emergence of a limit cycle regime for the coupled species $x,y$. Clearly, the regimes identified by means of equations (\ref{eq_ad}), which apply under the adiabatic approximation, reflect back on the original model constituted by equations (\ref{eq1}) and (\ref{eq2}). More concretely, the different zones as delineated with the reference model (\ref{eq_ad}), provide a reliable guide (data not shown) to anticipate the behavior of the system in its complete conception, as stipulated by eqs. (\ref{eq1}) and (\ref{eq2}). Focus for instance on the domain where the reduced model (\ref{eq_ad}) is predicted to display a limit cycle behaviour (dark blue region in Figure \ref{Figure1}). Then, for fixed $\gamma$ and large associated timescales, the initial model made of Eqs. (\ref{eq1}) and (\ref{eq2}) should also display periodic oscillations for the fraction of active excitatory ($x$) and inhibitory ($y$) neurons. A notable consequence is that  variables $R_E$ and $R_I$, coupled to the former, can in principle oscillate, an interesting dynamical regime which cannot be reproduced in the realm of the standard WC model. In the top panel of Figure \ref{Figure4}, the dynamical evolution of $R_x$ and $R_y$, as follows direct integration of equations  (\ref{eq1}) and (\ref{eq2}),  is plotted for different values of $\tau_R=\tau_{RE}=\tau_{RI}$ and  $\tau_D=\tau_{DE}=\tau_{DI}$, for fixed $\gamma$.

Interestingly, the oscillations for both $R_x$ and $R_y$ are present, but they fade away when increasing $\tau_R$ (and thus $\tau_D$). Notice that, as predicted by the adiabatic approximation, species $x$ and $y$ consistently oscillate, independently of the imposed time scales, at constant $\gamma$. The above reported behavior can be rationalized, as follows. Focus on the coupled pair $x$ and $R_E$ (similar considerations apply to  $y$ and $R_I$); assume that the latter quantities can be approximated by an harmonic ansatz of the type:

\begin{eqnarray}
    x(t)=A_xe^{i\Omega_xt+i\phi_x}+\bar{x} 
    \label{x_oscillator}\\
    R_E(t)=A_Ee^{i\Omega_Et+i\phi_E}+\bar{R_E} \nonumber
\end{eqnarray}

\begin{figure}
    \centering
    \begin{minipage}{0.4\textwidth}
    \centering
    \includegraphics[width=1\linewidth]{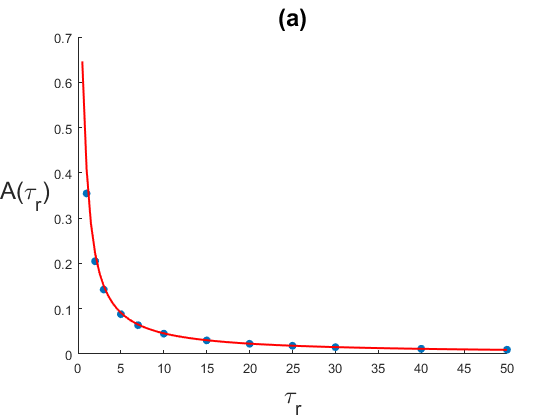}
    \end{minipage}
    \begin{minipage}{0.4\textwidth}
    \centering
    \includegraphics[width=1\linewidth]{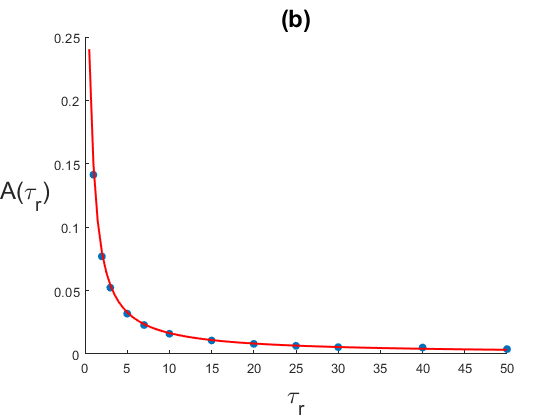}
    \end{minipage}
    \caption{$A_E$ versus $\tau_R$  for different values of $\gamma$, respectively equal to 2 (left panel) and 7 (right panel). Here, $a_E=a_I=1$, $h_E=-0.4$, $h_I=-0.5$, $I_E=0.5$, $I_I=0$,$\omega_{EE}=\omega_{IE}=1$, $\omega_{EI}=1.5$, $\omega_{II}=0.1$, $\xi_E=4.5$, $\xi_I=4$. For that choice of the parameters for $\gamma=2$ we numerically estimate $\omega=2.3$, $\bar{R_E}=4.1$, $A_x=0.5$ and $\bar{x}=0.2$. Instead, for $\gamma=7$, we have $\Omega=2.2$, $\bar{R_E}=4.4$, $A_x=0.6$ and $\bar{x}=0.2$.}
    \label{Figure5}
\end{figure}

Numerical insight suggests that the following assumptions holds, $\Omega_x=\Omega_E=\Omega$ (when, for large $\tau_R$,  $R_E$ converges monotonically to equilibrium, tiny oscillation are visible, of almost zero amplitude) and  $T_x=T_E=T=\frac{2\pi}{\omega}$. Then plugging Eqs.  
(\ref{x_oscillator}) in the first of Eqs. (\ref{eq2}), 
 multiplying both sides for $e^{i\Omega t+i\phi}$ and performing time average over a period $T$ yields:

\begin{equation}
    A_E(i\Omega \tau_r+1+\frac{\bar{x}}{\gamma})=-\frac{\bar{R_EA_x}}{\gamma}e^{-i\phi}
\end{equation}
and separating real and imaginary parts:

\begin{eqnarray}
    \Re(A_E)=\frac{\bar{R}_EA_x}{\gamma}\frac{(\Omega \tau_R \sin(\phi)-(1+\frac{\bar{x}}{\gamma})\cos(\phi))}{((1+\frac{\bar{x}}{\gamma})^2+\Omega^2\tau_R^2)} 
    \label{Amplitude real part}
    \\
     \Im{(A_E)}=\frac{\bar{R}_E A_x}{\gamma}\frac{(\Omega \tau_R \cos(\phi)+(1+\frac{\bar{x}}{\gamma})\sin(\phi))}{((1+\frac{\bar{x}}{\gamma})^2+\Omega^2\tau_R^2)}
\end{eqnarray}

from which one can estimate the amplitude of the oscillations \\ $A_E = \sqrt{\Re(A_E)^2 + \Im(A_E)^2}$, as a function of the time scale $\tau_R$, for different choices of $\gamma$. The other parameters are estimated from direct numerical experiments, carried out for the complete model made of Eqs. (\ref{eq1}) and (\ref{eq2}). In Figure \ref{Figure5} the prediction based on the obtained formula for $A_E$ vs. $\tau_R$ (solid line) is confronted with the values gathered from numerical integration of the model, for different values of $\gamma$. The found agreement, as it can be appreciated by visual inspection,  corroborates a posteriori the hypothesis put forward in the analysis.

{\bf Conclusions.} We have here proposed a generalized version of the celebrated Wilson-Cowan (WC) model which accounts for the self-consistent evolution of the synaptic resources. The model consists therefore of four coupled ordinary differential equations, which can be drastically simplified by adiabatically eliminating $R_E$ and $R_I$. We are therefore left with two equations for the fraction of active excitatory and inhibitory  neurons which present extra non linear terms, as compared to the WC benchmark model. This latter equations enable us to draw the phase diagram in the relevant parameter planes, by returning a general picture that applies also to the model in its original conception. Among the peculiar dynamical regimes, which can be accessed within the proposed schemes, worth mentioning are the oscillations for the resources concentration, which reflect the predicted limit cycle for species $x$ and $y$. In conclusion, the postulated frameworks, including the reduced model obtained under adiabatic elimination, provide an interesting generalization on current computation neuroscience models. As such, they can help to shed further light on a large gallery of relevant dynamical regimes which cannot be challenged under the customary WC descriptive picture. 

\bibliographystyle{unsrt}
\bibliography{bib_tommaso}

\section*{Acknowledgments}    
The authors would like to acknowledge the support by \#NEXTGENERATIONEU (NGEU) funded by the Ministry of University and Research (MUR), National Recovery and Resilience Plan (NRRP), project MNESYS (PE0000006)—A multiscale integrated approach to the study of the nervous system in health and disease (DN. 1553 11.10.2022).

\end{document}